\documentclass[aps,  showpacs]{revtex4}

\begin{document}
\title{Comment on 'Relativistic electron wavepackets carrying angular momentum'}
\author{S. C. Tiwari \\
Department of Physics, Institute of Science,  Banaras Hindu University,  and Institute of Natural Philosophy, \\
Varanasi 221005, India }
\begin{abstract}
This is a comment on arXiv:1611.04445 (PRL, 118, 114801 (2017)).
It is pointed out that the fundamental problems in light beam vortices and the relativistic electron vortices are not identical and have subtle differences. The significance of two length scales is underlined for the electron vortices. Local gauge transformation on the Gordon current admits vortex structure.
\end{abstract}
\maketitle
 
 Electron beams in free space having energy of 100 kev were experimentally demonstrated to have phase singularities carrying orbital angular momentum (OAM) by Uchida and Tonomura in 2010 \cite{1}. Since then many experiments have generated relativistic electron vortices in the beam energy range of 100-300 kev. Theoretical understanding of this phenomenon is essentially based on two ideas: analogy with the optical vortices and the well-known vortex solutions of nonrelativistic Schroedinger wave equation for electron. A recent paper \cite{2} argues that the origin of the electron vortices must be sought using Dirac equation. Authors construct solutions of the Dirac equation in an ingenious approach using the solutions of the Klein-Gordon equation. A detailed analysis shows that the relationship between OAM and the nature of vortices differs markedly from that of the nonrelativistic treatment. One of the striking results concerns the absence of the vortex line singularity in the Dirac current. Authors make no comment on the assumed analogy to the optical vortices in the literature. Curiously in another important contribution on this subject \cite{3} the significance of this analogy is underlined stating that, 'We are encouraged by the fact that analogous issues arose and were resolved for light, which is the quintessential relativistic field'.
 
The aim of the present note is to critically examine three fundamental issues. First one concerns the unsatisfactory use of the words electron and photon for the observed phenomena on electron beams and light beams respectively. Maxwell field equations describe macroscopic electromagnetic fields, while Schroedinger wave equation and Dirac equation describe single particle in the orthodox interpretation \cite{4}. For an insightful discussion on the macroscopic rendition of wavefunction we refer to Chapter 21 in \cite{5}.

The second question is regarding the physical nature of the problem in decomposing the total angular momentum (TAM) into spin (SAM) and OAM parts. Recall that in a relativistic field theory the canonical TAM tensor and the one constructed from the symmetric energy-momentum tensor are both divergenceless but differ by a pure divergence term \cite{6}. In the case of the electromagnetic field the lack of the gauge invariance for SAM and OAM parts, and the rest mass zero for photon in the quantized theory are the main issues. For the relativistic electron both issues are nonexistent, instead the main problem common with the electromagnetic field is that SAM and OAM are not conserved separately. Dirac considers the problem of electron in a spherically symmetric potential to establish the conservation for TAM \cite{7}. Thus the nature of the problem for light beams \cite{8} and electron beams is different in subtle ways.

Authors \cite{2} besides quoting Dirac \cite{7} make an important observation: the role of an intrinsic length scale and assume it to be the Compton wavelength $\lambda_c =\frac{\hbar}{mc}$. The nature of vortex lines is discussed using Dirac current and its Gordon decomposition. In the last paragraph of their paper a significant question is raised casting doubt on the existence of the electron vortices. I think the authors have missed two crucial physical ingredients in their arguments. Firstly there are two characteristic length scales not one for the electron, namely $\lambda_c$ and electron charge radius $r_e =\frac{e^2}{m c^2}$. Velocity of light appears in both, however though the Planck constant does not appear in $r_e$ it is smaller than $\lambda_c$ by a factor of fine structure constant $\alpha =\frac{e^2}{\hbar c}$. A thorough discussion on zitterbewegung, charge and mass centers, and the interpretation of Gordon current \cite{9} is necessary to have deeper understanding on the physics of vortices. Authors do not even mention this. Could there be a sub-quantum origin \cite{10} of the electron charge radius? It is remarkable that $\frac{e^2}{c}$ has the dimension of angular momentum. Even if speculations made in \cite{9,10} are ignored the issue of two length scales deserves attention for the sake of completeness of the analysis given in \cite{2,3}.

Another physical argument is that of local gauge transformation of the currents in the Dirac theory under which
\begin{equation}
\Psi ~ \rightarrow ~ e^{i \alpha \theta} \Psi
\end{equation}
The Dirac current, and the spin current in the Gordon decomposition are invariant under (1). However the Gordon current transforms to
\begin{equation}
J^\mu_G =\frac{i e \hbar}{2mc} [\bar{\Psi} \partial^\mu \Psi -(\partial^\mu \bar{\Psi}) \Psi] - e r_e \bar{\Psi} \Psi \partial^\mu \theta
\end{equation}
Following the approach given in \cite{2} the presence of the last term for a nontrivial singular phase factor $\theta$ leads to a vortex line. Therefore the electron vortex does exist. Of course, experimentally the vortex core region becomes important that unfortunately was not measured in \cite{1}. Note that here we are not interested in quantum electrodynamics or the so called longitudinal photon theory.
The factor $\alpha$ in phase (1) is introduced to make Eq.(2) consistent with the expression for the Gordon current in the presence of the electromagnetic interaction.

\end{document}